\documentclass[aps,prb,twocolumn,showpacs,floatfix,superscriptaddress,
citeautoscript
,cite]{revtex4}

\usepackage{graphicx} % Include figure files
\usepackage[utf8]{inputenc}
\usepackage[T1]{fontenc}
\usepackage{lmodern}
\usepackage{color}

\begin{document}

\author{ S. Horzum}
	\email{seyda.horzumsahin@ua.ac.be}
\affiliation{Departement Fysica, Universiteit Antwerpen, Groenenborgerlaan 171,
B-2020 Antwerpen, Belgium}
\affiliation{Department of Engineering Physics, Faculty of Engineering, Ankara
University, 06100 Ankara, Turkey}

\author{H.  Sahin}
	\email{hasan.sahin@ua.ac.be}
\affiliation{Departement Fysica, Universiteit Antwerpen, Groenenborgerlaan 171,
B-2020 Antwerpen, Belgium}

\author{S. Cahangirov}
	\email{seycah@gmail.com}
\affiliation{Nano-Bio Spectroscopy group, Dpto.~F\'isica de Materiales, Universidad del Pa\'is Vasco, Centro de F\'isica de Materiales CSIC-UPV/EHU-MPC and DIPC, Av.~Tolosa 72, E-20018 San Sebasti\'an, Spain}

\author{P. Cudazzo}
	\email{pierluigi.cudazzo@ehu.es}
\affiliation{Nano-Bio Spectroscopy group, Dpto.~F\'isica de Materiales, Universidad del Pa\'is Vasco, Centro de F\'isica de Materiales CSIC-UPV/EHU-MPC and DIPC, Av.~Tolosa 72, E-20018 San Sebasti\'an, Spain}

\author{A. Rubio}
	\email{angel.rubio@ehu.es}
\affiliation{Nano-Bio Spectroscopy group, Dpto.~F\'isica de Materiales, Universidad del Pa\'is Vasco, Centro de F\'isica de Materiales CSIC-UPV/EHU-MPC and DIPC, Av.~Tolosa 72, E-20018 San Sebasti\'an, Spain}

\author{T. Serin}
	\email{tulay.serin@eng.ankara.edu.tr }
\affiliation{Department of Engineering Physics, Faculty of Engineering, Ankara
University, 06100 Ankara, Turkey}

\author{F. M. Peeters}
	\email{francois.peeters@ua.ac.be}
\affiliation{Departement Fysica, Universiteit Antwerpen, Groenenborgerlaan 171,
B-2020 Antwerpen, Belgium}

\title{Phonon Softening and Direct to Indirect Bandgap Crossover in Strained
Single Layer MoSe$_{2}$}
\date{\today}

\begin{abstract}

Motivated by recent experimental observations of Tongay \textit{et al.} [Tongay
\textit{et al.}, Nano Letters, \textbf{12}(11), 5576 (2012)] we show how the
electronic properties and Raman characteristics of single layer MoSe$_{2}$ are
affected by elastic biaxial strain. We found that with increasing strain: (1)
the $E^{\prime}$ and $E^{\prime\prime}$ Raman peaks ($E_{1g}$ and $E_{2g}$ in
bulk) exhibit significant red shifts (up to $\sim$30 cm$^{-1}$), (2) the
position of the $A_{1}^{\prime}$ peak remains at 180 cm$^{-1}$ ($A_{1g}$ in
bulk) and does not change considerably with further strain, (3) the dispersion
of low energy flexural phonons crosses over from quadratic to linear and (4) the
electronic band structure undergoes a direct to indirect bandgap crossover under
$\sim$3$\%$ biaxial tensile strain. Thus the application of strain appears to be
a promising approach for a rapid and reversible tuning of the electronic,
vibrational and optical properties of single layer MoSe$_{2}$ and similar
MX$_{2}$ dichalcogenides.

\end{abstract}

\maketitle

\section{Introduction}

The discovery of graphene\cite{novo, geim} is expected to play an important role
in future nanoscience and nanotechnology applications. Recent advances in
nanoscale growth and mechanical exfoliation techniques have led not only to the
fabrication of high-quality graphenes but also the emergence of several new
classes of two-dimensional (2D) structures such as ultra-thin transition metal
dichalcogenides (TMDs). Just like graphene TMDs have hexagonal crystal
structure composed of layers of metal atoms (M) sandwiched between layers of
chalcogen atoms (X) with stoichiometry MX$_{2}$.
TMDs have electronic semiconducting or metallic properties and most of bulk
TMDs possess indirect band gaps of the order 1-2 eV \cite{gap}. The synthesis of
several TMDs has been realized experimentally \cite{science 2011, prb-2002,
novo-pnas-2005, nature-nano} and the stability and electronic properties of
various single layer dichalcogenides was recently reported.\cite{mx2} Recent
studies of several semiconducting TMDs such as MoS$_{2}$, WS$_{2}$,
MoTe$_{2}$ and MoSe$_{2}$ have shown that the band gap increases and transforms
to a direct band gap with decreasing number of layers.\cite{nature-nano,PRL-105,
prb84, apl99} These sizeable band gaps make them well suited for electronic
applications such as transistors, photodetectors and electroluminescent devices.
Another unique feature of two-dimensional ultra-thin materials is the
possibility to apply large reversible elastic strain. It has been shown that
graphene can be strained up to 20 $\%$ of its ideal structure with only small
changes in its electronic band structure, which is in contrast to
TMDs.\cite{Bull.Mater,prb-2011,yun,sca}

The most recent efforts have been directed towards the synthesis and
manipulation of  molybdenum diselenide (MoSe$_{2}$) single layers. In
addition to preliminary reports on synthesis of few-layer MX$_{2}$
structures\cite{condens-mat,apl-sefa}, Tongay \textit{et al.}
demonstrated that a single layer of MoSe$_{2}$ possesses a direct optical gap of
1.55 eV and exhibits good thermal stability.\cite{sefamose2} However, to our
knowledge, no research exists addressing the question of how electronic
properties and lattice dynamics of single layer MoSe$_{2}$ are affected by
strain. In the present work, we investigate the effect of biaxial strain on the
electronic and vibrational properties of single layer MoSe$_{2}$ structures.
Our calculations revealed that applying biaxial strain is able to tune the
Raman characteristics and electronic band structure of single
layer MoSe$_{2}$. We expect that this study will offer new opportunities for
strain-engineered nanoscale optoelectronic device applications.

\section{Computational METHODOLOGY}

Calculations of physical properties of equilibrium and strained structures were
carried out in the framework of density functional theory (DFT), using the
plane-wave self-consistent field (PWSCF) code as implemented in the
QUANTUM-ESPRESSO package \cite{quantumespresso}. The generalized gradient
approximation (GGA) of Perdew-Burke-Ernzerhof (PBE) was used for the
exchange-correlation potential \cite{pbe}. A plane-wave kinetic energy cutoff of
40 Ry and density cutoff of 400 Ry were used. Brillouin zone integration was
performed using a shifted 27$\times$27$\times$1 Monkhorst-Pack
mesh\cite{kpoint}. To
eliminate the interaction emerging from periodic boundary conditions
calculations are performed with a large unitcell including 12~{\AA} vacuum space
between adjacent MoSe$_{2}$ single layers. Ground states and total energies of
all systems were obtained after full geometry relaxation with forces on the
atoms smaller than 0.02 eV/{\AA}. Phonon frequencies and phonon eigenvectors
were calculated in a 4$\times$4$\times$1 \textbf{q}-grid using the density
functional perturbation theory (DFPT).\cite{dfpt}

\section{results and Discussion}

\subsection{Single Layer MoSe$_{2}$}

\begin{figure}
\includegraphics[width=8.5cm]{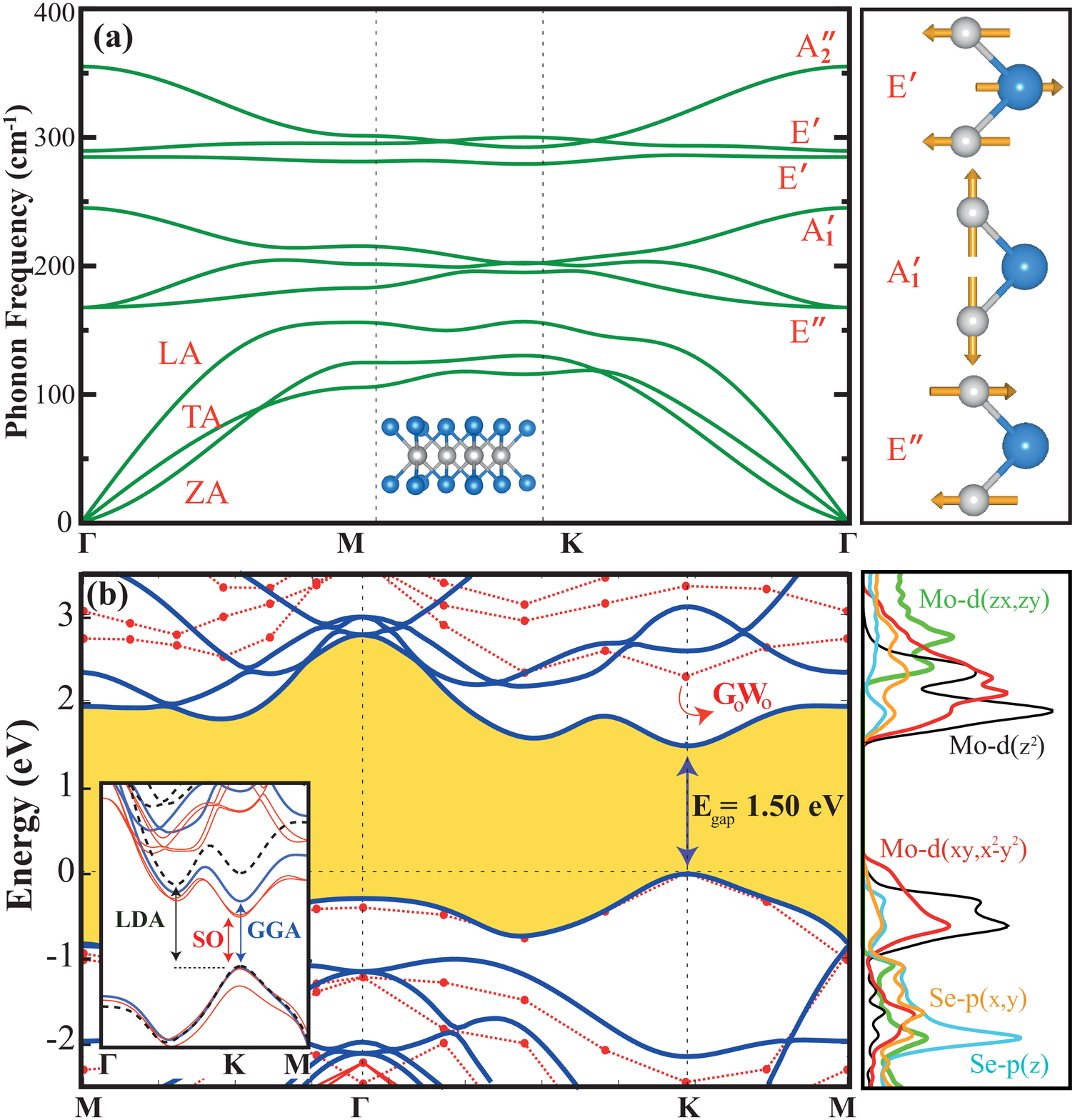}
\caption{\label{fig1}
(Color online) (a) Phonon dispersion for single layer MoSe$_{2}$ and
displacement directions of Raman-active modes. Inset: top view of atomic
structure of MoSe$_{2}$ (b) Electronic band structure and
partial density of states calculated within GGA and $G_0W_0$. The energies are
relative to the Fermi level (i.e. $E_{\mathrm{F}}=0$) Inset: Comparison of the
band edges calculated by GGA, GGA+SO and LDA.}
\end{figure}

%\subsubsection{Lattice Dynamics}

Atomically thin MoSe$_{2}$ is the most recently synthesized member of the
ultra-thin transition metal dichalcogenides. A single layer structure of
MoSe$_{2}$ can be viewed as a 3-layer stacking of Mo and Se atoms wherein
molybdenum atoms are sandwiched between layers of trigonally arranged
selenium atoms. In this configuration, known as 1H structure, each Mo atom is
coordinated to six Se atoms. In the equilibrium geometry of the single layer
MoSe$_{2}$, the Mo-Se distance, the Se-Se distance and the lattice constant of
the
hexagonal unit cell are calculated to be 2.528, 3.293 and |\textbf{a}|=3.321 \AA,
respectively. Thus, in comparison with single layer MoS$_{2}$ having lattice
constant 3.12 \AA, MoSe$_{2}$ has a larger atomic structure due to the larger
atomic radius of the Se atoms. In Fig. \ref{fig1}(a) we present the calculated
phonon dispersion and vibrational characteristics of R-active modes of single
layer MoSe$_{2}$. It is seen that the frequencies of all phonon branches in the
whole Brillouin Zone have positive values i.e., MoSe$_{2}$ crystal can remain
stable by generating the required restoring force against atomic distortions.
Although the synthesis of single layers of MoSe$_{2}$ have been achieved only on
various substrates so far, energy optimization and phonon dispersion
calculations show that freestanding MoSe$_{2}$ monolayers are quite stable.

%\subsubsection{Electronic Structure}

The calculated band structure shown in Fig. \ref{fig1}(b) shows that the single
layer MoSe$_{2}$ is a semiconductor with a direct bandgap at the K
high symmetry point. The top of the valence band is mainly composed of
Mo-$d_{(x^{2}-y^{2})}$ orbitals and, albeit small, also by Mo-$d_{(xy,yz,zx)}$
and Se-$p_{(x,y)}$ orbitals. Though the electrons occupying 4$d$ and 5$s$ shells
of a neutral Mo atom are treated as valence electrons, 5$s$ states have a
negligible contribution even away from the Fermi level. However, it is seen that
all Mo-$d$, Se-$p_{(x,y)}$ and Se-$p_{(z)}$ orbitals are hybridized to form the
conduction band edge. Note that $x$ and $y$ component of the orbitals have the
same energy due to the two-dimensional lattice symmetry. Due to the
presence of another minimum in the conduction band edge between the $\Gamma$ and
K points we also used local density approximation (LDA) for the exchange
correlation functional. Tongay \textit{et al.}\cite{sefamose2} demonstrated that
such a neighboring conduction band dip is quite sensitive to temperature-induced
strain and layer-layer interaction. The lattice constant of a single-layer
MoSe$_{2}$ is found to be smaller (3.22 \AA) in LDA because of the
overestimation of the strength of the covalent bonds. The inset of Fig.
\ref{fig1}(b) shows a zoom of both LDA, GGA and GGA+SO band structures along the
$\Gamma$-K-M path in the Brillouin zone. We see a splitting of the band edges at
the vicinity of K point when including spin-orbit (SO) interactions.
Interestingly, LDA predicts that single layer MoSe$_{2}$ is an indirect
semiconductor where the top of the valence band is at the K point, and the
bottom of the conduction band is found to be between the $\Gamma$ and K points.
It appears that the physically correct description of single layer MoSe$_{2}$
structure that was reported experimentally as a semiconductor\cite{sefamose2}
with a direct bandgap is achieved by using the GGA functionals. Therefore, in
the rest of our study PBE calculations will be employed.

Although our PBE calculation captures the qualitative nature of the band edge
states
it underestimates the band gap. To correct the band gap we use the $G_0W_0$
approximation\cite{hedin,shishkin,onida} as implemented in the VASP
package.\cite{kresse,paw} It was shown before that this kind of
calculation has to be converged with respect to the vacuum spacing.\cite{vacuum,wirtz}
We calculate the quasi-particle shifts using (12$\times$12$\times$1) k-point
grid with vacuum spacings of 15~\AA, 20~\AA, 25~\AA, and 30~\AA~and then
extrapolate them to infinity in a similar fashion as
done in Ref. [~\onlinecite{vacuum}].
Our $G_0W_0$ calculation results in a band gap value of 2.33~eV. 
Note that, excitonic effects are pronounced in 2D materials as
MoSe$_{2}$ and we need to calculate the binding energy of the exciton in order
to estimate the optical gap.\cite{wirtz,exc-mx2,qp-mos2,bse-mose2,cud-prl,cud-prb} 
Due to the parabolic shape of the conduction and valence bands we assume that
the exciton has the Mott-Wannier character and the effective mass approximation
can be used. Hole and electron effective masses calculated from the band
structure are $m_{h}^{*}$ = 0.65 and $m_{e}^{*}$ = 0.53 (in units of
electron mass). Moreover, we account for the 2D polarizability of the
system using
the model introduced by Cudazzo \textit{et al}.\cite{cud-prb} Using parameters
reported by Ref. [~\onlinecite{exc-mx2}] we find the binding energy of the
exciton 0.7~eV, which is in agreement with the value obtained by solving the
Bethe-Salpeter equation.\cite{exc-mx2,bse-mose2} Combining this with our result
from $G_0W_0$ calculation we estimate the optical gap to be 1.63~eV, which is
close to the reported experimental value of 1.56 eV.

\begin{figure}
\includegraphics[width=8.5cm]{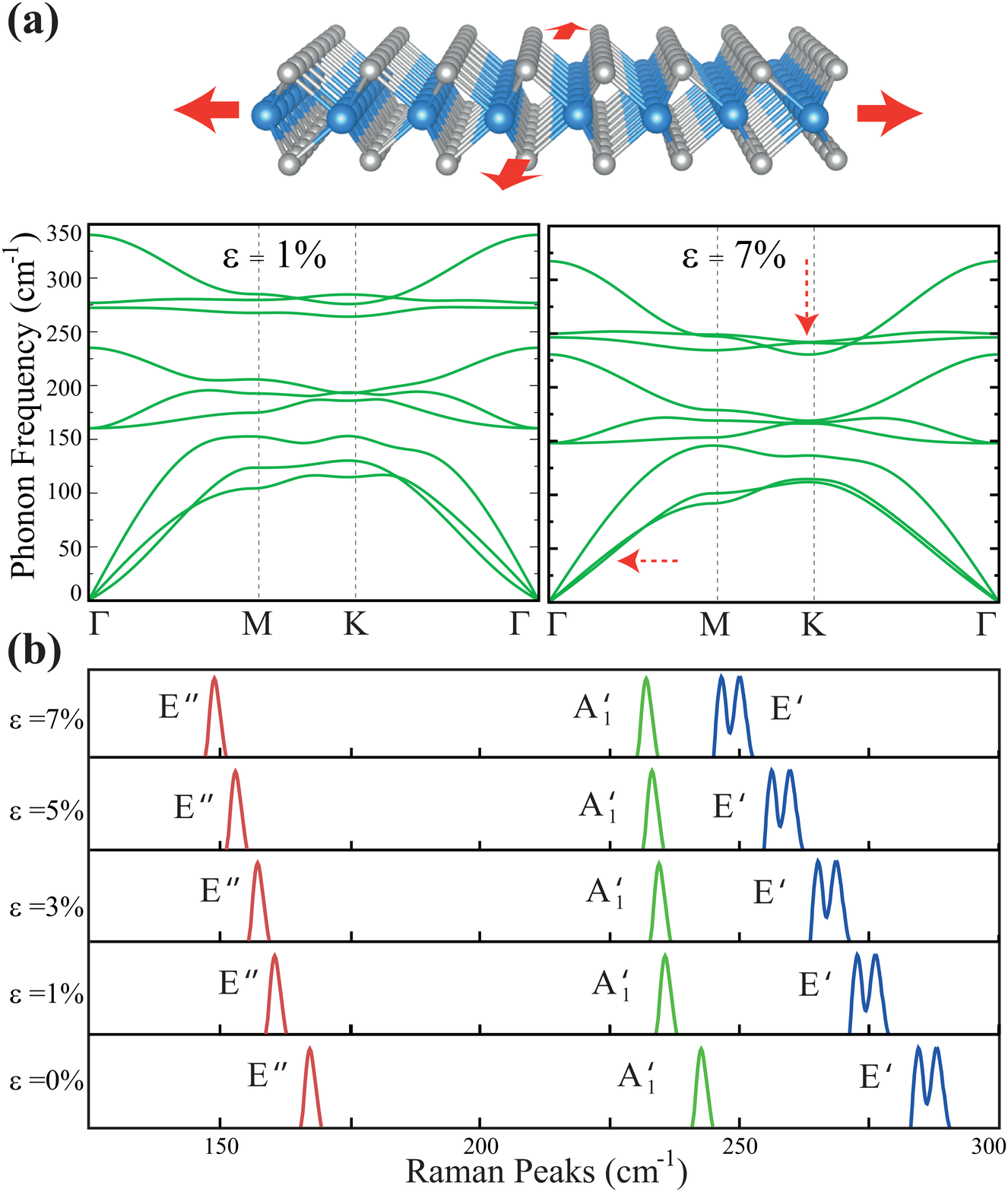}
\caption{\label{fig2}
(Color online) (a) Phonon dispersions for 1$\%$ and 7$\%$ strained single layer
MoSe$_{2}$ (b) Evolution of the Raman-active modes as a function of applied
strain.}
\end{figure}

\subsection{Lattice Dynamics and Phonon Softening}

While bulk MoSe$_{2}$ crystal has the D$_{6h}$ point group symmetry, it
turns into D$_{3h}$ for single layer structure. Just
like in MoS$_{2}$,\cite{phononmos2} lattice vibrations of single
layer MoSe$_{2}$ is characterized by nine phonon branches including three
acoustic and six optical branches. Analysis of lattice dynamics shows that the
decomposition of the vibration representation at the $\Gamma$ point is $\Gamma =
A_{1}^{\prime} + 4E^{\prime} + 2E^{\prime\prime} + 2A_{2}^{\prime\prime} $.
While LA and TA acoustic branches have linear dispersion, the frequency of the
out of plane flexural (ZA) mode has a quadratic dispersion in the vicinity of
q=0. Near to the $\Gamma$ point the in-plane sound velocity of the LA and TA
modes are found to be 1.5$\times$10$^{3}$ and 0.9$\times$10$^{3}$ m/s,
respectively. It is also seen that differing from graphene's phonon dispersion
the acoustic and optical modes are well-separated from each other. Differing
from MoS$_{2}$ and WS$_{2}$ structures in which the $A_{1}^{\prime}$ mode is the
highest Raman-active mode, in MoSe$_{2}$ the $A_{1}^{\prime}$ mode is located in
between the $E^{\prime}$ and $E^{\prime\prime}$ modes. In Fig. \ref{fig1}(a)
the atomic displacements of the R-active modes are presented. Interestingly,
although recent Raman spectroscopy measurements of Tongay \textit{et
al.}\cite{sefamose2} revealed that only one characteristic Raman peak of
single-layer MoSe$_{2}$ $E_{2g}$ is observable at 240 cm$^{-1}$, our symmetry
analysis predicts the existence of two more Raman-active modes. Significantly
decreased intensity of Raman peak around 290 cm$^{-1}$ can be explained by
considering the coupling of in-plane counter-phase vibration of MoSe$_{2}$ with
the SiO$_{2}$/Si substrate. Moreover, in experiment the disappearance of
$E^{\prime\prime}$ ($E_{1g}$ in bulk) Raman shift around 170 cm$^{-1}$ is due to
the inactivity of this vibrational mode to normally incident light.

Recent experiments on graphene and other few-layer materials have revealed that
the application of biaxial and uniaxial tensile stress is possible
by using flexible substrates. We will investigate now how the lattice dynamics
of single layer MoSe$_{2}$ is affected by biaxial strain. 

Fig. \ref{fig2} shows phonon
dispersion curves and the evolution of Raman peaks as a function of applied
strain. Below 1\% strain, the $E^{\prime}$ branch of single layer MoSe$_{2}$
experiences $\sim$10 cm$^{-1}$ softening which is larger than
that of the $E_{g}$ mode (diatomic corresponding to $E^{\prime}$) of
graphene.\cite{moh,hua,zab} It is seen that the low-energy flexural phonons
(ZA) of unstrained
single layer MoSe$_{2}$ turns almost in a linear dependence under 7$\%$ strain.
Such a change in phonon dispersion implies a significant decrease in
scattering of electrons by the flexural phonons and an increase in electron
mobility for strained MoSe$_{2}$. For the $E^{\prime}$ mode there is a small
splitting at the $\Gamma$ point, as in slightly polar materials such as
MoS$_{2}$ and WS$_{2}$.\cite{phononmos2} It is also seen that due to the
weakening of the covalent bonds under strain the $E^{\prime}$ and
$E^{\prime\prime}$ modes (that correspond to the $E_{1g}$ and $E_{2g}$ in bulk
MoSe$_{2}$, respectively) exhibit significant red shifts of about $\sim$30
cm$^{-1}$. Especially, at the K point, the highest optical mode dips even below
the band corresponding to the $E^{\prime}$ mode. Another remarkable point is the
$A_{1}^{\prime}$ mode (corresponding to the $A_{1g}$ in bulk) that vibrates
perpendicular to the applied strain which experiences a smaller softening and
is almost pinned to 231 cm$^{-1}$ for strain values higher than 1$\%$.
Similar direction-dependent softening behavior was observed for various
types of materials.\cite{C60,lan}

\subsection{Strain-Induced Direct-to-Indirect Bandgap Crossover}

\begin{figure}
\includegraphics[width=8.5cm]{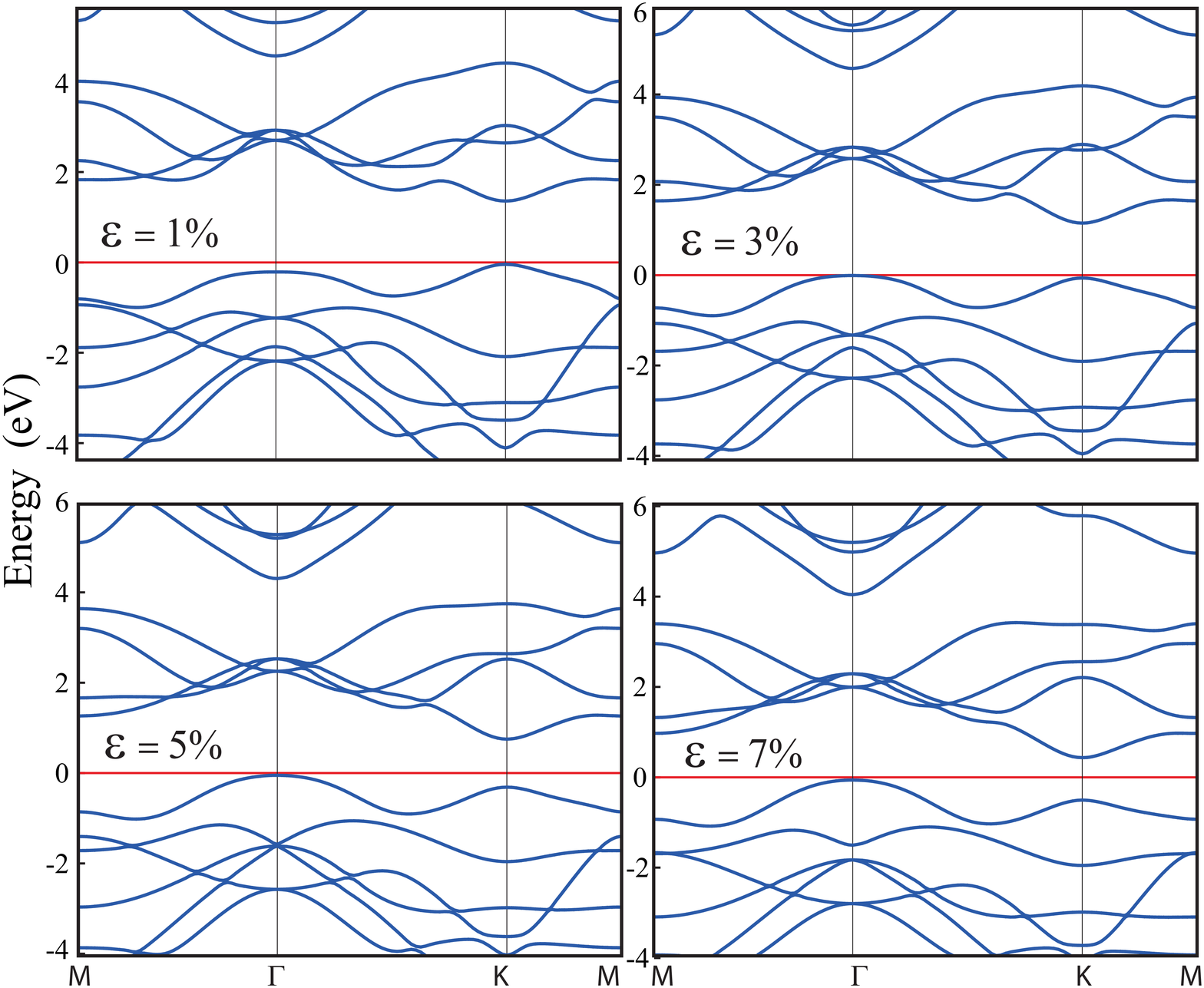}
\caption{\label{fig3}
(Color online) Evolution of the electronic band dispersion of single layer
MoSe$_{2}$ as a function of strain. Fermi level shown by red line is set to
zero in the energy spectrum.}
\end{figure}

In addition to strain-induced changes in the lattice dynamics, the electronic
properties of single layer MoSe$_{2}$ may exhibit surprising deviations.
Existence of a bandgap even in the monolayer structure shows strong
bonding nature and the covalent character of MoSe$_{2}$ layers. Though the
covalent bonds formed from valance electrons of Mo and Se atoms are not broken
by elastic strain, the strength of the bond is weaken with the increasing
distance.
Therefore band energies and dispersion can be expected to be strongly
dependent on tensile strain.
Evolution of the electronic band dispersion under finite biaxal strain
($\varepsilon = 1-7\%$) is shown in Fig. \ref{fig3}. It is clearly seen that
the band edges at the $\Gamma$ and K symmetry points are influenced
significantly by biaxial tensile strain. Up to 3$\%$ strain, single layer
MoSe$_{2}$ remains a direct bandgap semiconductor with both the top of the
valence band and the bottom of the conduction band located at the K point.
However, when $\sim$3$\%$ strain is applied, as a result of the band shifting at
the K point, two valence band edges with the same energy appear in the Brillouin
Zone while the conduction band edge remains fixed in its original position.
Here, with $\sim$3$\%$ strain one has both a direct (K$\rightarrow$K) and an
indirect ($\Gamma \rightarrow$K) bandgap.

\begin{figure}
\includegraphics[width=8.5cm]{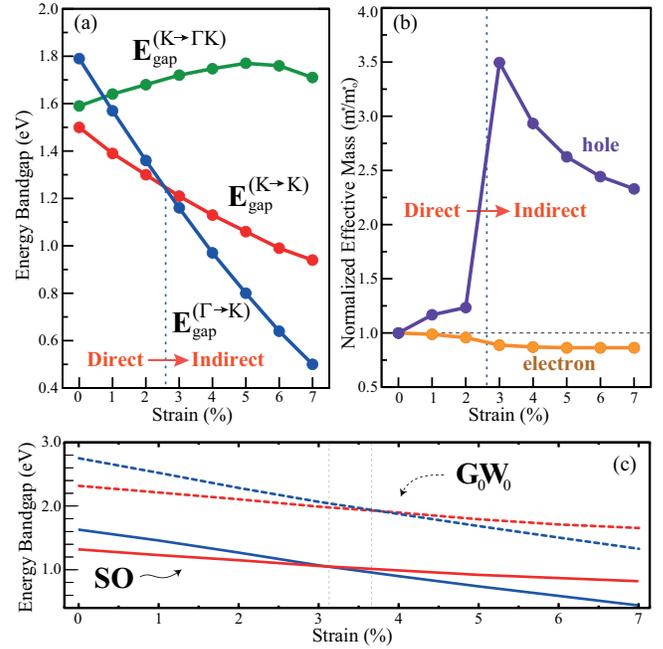}
\caption{\label{fig4}
(Color online) Evolution of (a) direct and indirect energy band gaps, (b)
electron and hole effective mass, normalized to unstrained value
($m_{e}^{*}/m_{e0}^{*}$ and $m_{h}^{*}/m_{h0}^{*}$ ), under biaxial
tensile strain. (c) Energy bandgaps calculated by using PBE+SO and
PBE+$G_{0}W_{0}$ are shown by solid and dashed lines, respectively. Red, green
and blue colors are used for transitions
K$\rightarrow$K, K$\rightarrow\Gamma$K and $\Gamma\rightarrow$K, respectively.}
\end{figure}

Such an interesting electronic structure may result in the presence of
two types of holes having different effective masses and the coexistence of
direct and indirect band transitions in the optical spectra. We also found that
increasing further the strain results in the separation (and shifting) of
the uppermost valence band towards the conduction band accompanied by the
lowering of the conduction band dip at the K point. Moreover, the second
conduction band edge located (between $\Gamma$ and K) very close to the dip of
the conduction band at the K point moves upwards in energy. In addition,
band edges of MoS$_{2}$ also experiences similar effects with increasing
tensile strain.\cite{yun,sca} The evolution of direct (K$\rightarrow$K)
and indirect (K$\rightarrow \Gamma$K and $\Gamma \rightarrow$K) energy
bandgaps under tensile stress are presented in Fig. \ref{fig4}(a). 
Both direct and indirect bandgaps decrease monotonically with increasing strain, 
however the rate of decrease is faster for the indirect bandgap. 
Note that, the variation of band gaps with increasing strain has the
same qualitative behavior for PBE, PBE+SO and PBE+$G_{0}W_{0}$ calculations. 
We predict that the direct-to-indirect crossover takes place at 3-4$\%$ strain.

As a consequence of the strain-dependent transition of the valence band edge
from the K to $\Gamma$ point the effective mass of the holes (and electrons)
propagating through the single layer MoSe$_{2}$ lattice changes dramatically.
From Fig. \ref{fig4}, it is seen that while the mass of the electrons at the
band edge does not change notably, the hole mass shows significant deviations
under strain. Since the valence band edge is sharpened by biaxial strain the
effective mass of the holes is decreased and after the bandgap crossover holes
belong to the uppermost valence band at the $\Gamma$ point inversely affected by
increasing strain. Note that the holes under 7$\%$ strain are $\sim$2
times heavier than that for unstrained single layer MoSe$_{2}$.

%\section{conclusions}

In summary, motivated by the recent experimental study of Tongay \textit{et
al.}\cite{sefamose2} on single layer MoSe$_{2}$ we investigated the
electronic properties and lattice dynamics of single layer MoSe$_{2}$ as a
function of biaxial strain. We showed that GGA describes more
accurately the electronic structure of single layer MoSe$_{2}$ and that LDA
predicts the wrong nature of the bandgap. We found that monolayer MoSe$_{2}$ has
four R-active characteristic modes: among these, the modes having counter-phase
in-plane motion are significantly modified by strain and they soften by
increasing strain. Red-shift in the position of these resonant Raman peaks are
predicted to be $\sim$30 cm$^{-1}$. However the $A_{1}^{\prime}$ branch
(corresponds to the $A_{1g}$ in bulk) is negligibly influenced by strain.
Moreover the linear dispersion of flexural mode in strained-MoSe$_{2}$ implies a
reduced scattering of electrons from these ZA phonons. Our findings also
revealed that the electronic band structure of single layer MoSe$_{2}$ also
undergoes substantial changes under biaxial tensile strain. While the direct
bandgap linearly decreases up to $\sim$3$\%$ tensile strain, a direct to
indirect bandgap crossover occurs when strain is further increased. Moreover the
appearance of a positive curvature at the K point in the phonon spectra and
possible metallicity for higher strain values suggests superconductivity in
highly strained single layer MoSe$_{2}$. Our findings indicate that the bandgap
of single layer MoSe$_{2}$ can be tuned reversibly by biaxial strain and is able
to capture a broad range of the solar spectrum.  

\section{Acknowledgements}

This work was supported by the Flemish Science Foundation (FWO-Vl) and
the Methusalem programme of the Flemish government. Computational resources were
partially provided by TUBITAK ULAKBIM, High Performance and Grid Computing
Center (TR-Grid e-Infrastructure). H. S. is supported by a FWO Pegasus Marie
Curie Long Fellowship.

\end{document}